\documentclass{elsart}
\usepackage{epsfig}

\begin{document} 

\begin{frontmatter}

\title{Domain Growth in a Multivariable non Potential System}

\author{R. Gallego, M. San Miguel and R. Toral} 
\address{Instituto Mediterr\'{a}neo de Estudios Avanzados, IMEDEA (CSIC-UIB)\\ 
Campus Universitat Illes Balears, E-07071 Palma de Mallorca, Spain}
\begin{abstract} 
We present a study of dynamical scaling and domain growth in a  non potential 
system that models Rayleigh--B\'{e}nard convection in a rotating cell. In
$d=1$, dynamical scaling holds, but the non potential terms 
modify the characteristic growth law with a crossover from
logarithmic to linear in time. In $d=2$ the non potential terms prevent
coarsening for any value of  the angular rotation speed.
\end{abstract} 

\end{frontmatter}

\section{Introduction and model}

Over the last decades, a large effort has been devoted to study the coarsening
processes that drive a system back to the equilibrium state \cite{GSMS,BRAY}.
One of the main points of interest is to investigate  the existence of
dynamical scaling in the late stages of  the evolution. Briefly stated,
dynamical scaling means that there exists a single characteristic length,
$R(t)$,  such that the domain structure is independent of time (in a
statistical sense) when all the lengths are scaled by $R(t)$. In other words,
the system evolves in a self-similar manner.  It is known that for potential
dynamics, i.e., for systems whose dynamical evolution involves the minimization
of a potential (free energy) with two equivalent minima, and after a transient
time in which domains are formed, there appear well defined interfaces
separating the equivalent states. The subsequent dynamics is governed by
interface motion.  The mechanism for interface motion, and therefore the
characteristic length $R(t)$, strongly depend on the dimensionality, $d$, as
well as whether the (scalar) order parameter is conserved or not during the
dynamical evolution. For non conserved order parameter the results are as
follows: in $d=1$, the domain boundaries (hereafter to be called \emph{kinks})
move due to the interaction between them, leading to a logarithmic growth with
time of the characteristic length, $R(t)\sim\log t$. In $d\ge 2$ the mechanism
for domain growth is curvature driven and the characteristic length behaves as
$R(t) \sim t^{1/2}$. 

It is the purpose of this paper to study the influence of non potential
effects (those that can not be derived from the minimization of a potential
function) on the coarsening process and, namely, on the growth law of the
characteristic domain size as well as the  validity of the scaling description
of the dynamics. To this end,  we have used a theoretical model proposed in the
context of fluid dynamics by Busse and Heikes~\cite{Busse-Heikes} to which
spatial dependent terms of the simplest diffusive form have been added:
\begin{equation}\label{eq:modelo}
  \begin{array}{rcl}
  \partial_{t} A_{1} &=& \nabla^{2}A_{1}+A_{1}\, (1-A_{1}^{2}
  -(\eta + \delta)\, A_{2}^{2}-(\eta-\delta)\, A_{3}^{2})	\\
  \partial_{t} A_{2} &=& \nabla^{2}A_{2}+A_{2}\, (1-A_{2}^{2}
  -(\eta + \delta)\, A_{3}^{2}-(\eta-\delta)\, A_{1}^{2})	\\
  \partial_{t} A_{3} &=& \nabla^{2}A_{3}+A_{3}\, (1-A_{3}^{2}
  -(\eta + \delta)\, A_{1}^{2}-(\eta-\delta)\, A_{2}^{2})
  \end{array}
\end{equation}
This models aims to represent the appearance of convection rolls in a
Rayleigh-B\'enard fluid subject to Coriolis forces due to rotation.  $A_1$,
$A_2$ and $A_3$ are the (real) amplitudes of the convection rolls in three
different space directions oriented at $120^0$ from each other. The parameters
$\eta$ and $\delta$ are related to physical properties of the fluid. In
particular $\delta$ is linked to the rotation angular velocity in such a way
that $\delta =0$ means no rotation of the fluid.  We can split
(\ref{eq:modelo}) into potential and non potential contributions \cite{chile}:
\mbox{$\partial_{t}A_{i}=-\delta{\mathcal F}/\delta A_{i}+\delta\cdot f_i$}
$(i=1,2,3)$. Therefore when $\delta=0$, the system adopts a potential form
with a potential $\mathcal F$.  When $\delta\neq 0$, the dynamics is said to be
non potential or non variational.

The set of equations (\ref{eq:modelo}) admits two kinds of homogeneous
stationary solutions, namely: three ``roll'' solutions $(A_{i}=1,\ A_{j}=0,\
i\neq j)$ and one ``hexagon'' solution $(A_{1}=A_{2}=A_{3}=(1+2\eta
)^{-1/2})$.  When the angular velocity is smaller than some critical value, 
$\delta <\delta_c = \eta-1$, the rolls are the only stable solutions and the
dynamics leads, after a short transient time, to a situation in which there
exist well-defined interfaces connecting two roll states. When the angular
velocity is greater than the critical value, $\delta >\delta_c$, the system
switches to a time dependent dynamics known as the K\"uppers-Lortz (KL)
instability \cite{KL}. The KL instability introduces a chaotic dynamics that
prevents coarsening and the typical domain size $R(t)$ saturates to a finite
value \cite{CMT}. In this paper we show that for angular rotation velocities
smaller than the critical value (hence, far away from  the KL instability) the
system coarsens in $d=1$ but not in $d=2$. The difference  lays on the fact
that in $d=2$ coarsening is  stopped due to the existence of points where three
different front lines meet. The non potential dynamics makes
the front lines rotate around the vortices.

\section{Fronts and domain growth in $\mathbf{d=1}$}

As mentioned before, for small enough angular velocities, $\delta < \delta_c$,
there are stable kink solutions connecting two different roll states, say 
$A_i$, $A_j$, the third amplitude being zero everywhere. In the potential
case, $\delta=0$, a solitary kink does not move because of the symmetry of the
connected roll states. When $\delta \ne 0$ this symmetry is broken and the kink
moves at a constant velocity, $v(\delta)$.  This can be computed by means of a
perturbation analysis through a solvability condition. The resulting expression
at leading order in $\delta$ is \cite{GSMT}:
\begin{equation}\label{eq:vel}
  v(\delta)=\delta \: 
  \frac
  {
  \int_{-\infty}^{\infty}dx\,
  A_{i}^{0}A_{j}^{0}\, (A_{j}^{0}\,\,
  \partial_{x}A_{i}^{0}-A_{i}^{0}\,\,\partial_{x}A_{j}^{0})
  }
  {
  \int_{-\infty}^{\infty}dx
  \left[ (\partial_{x}A_{i}^{0})^{2}+(\partial_{x}A_{j}^{0})^{2} 
  \right]
  }
\end{equation}
where $A_{i}^{0}$, $A_{j}^{0}$ are the amplitudes of the stationary potential
problem. With the help of this expression it is possible to know not only the
magnitude of the velocity but also the direction of the motion which is related
to the sign of $v$. Out of the six possible types of kinks connecting
different roll states, three move to the right and three to the left.

\begin{figure}[h]
\begin{center}
  \epsfig{figure=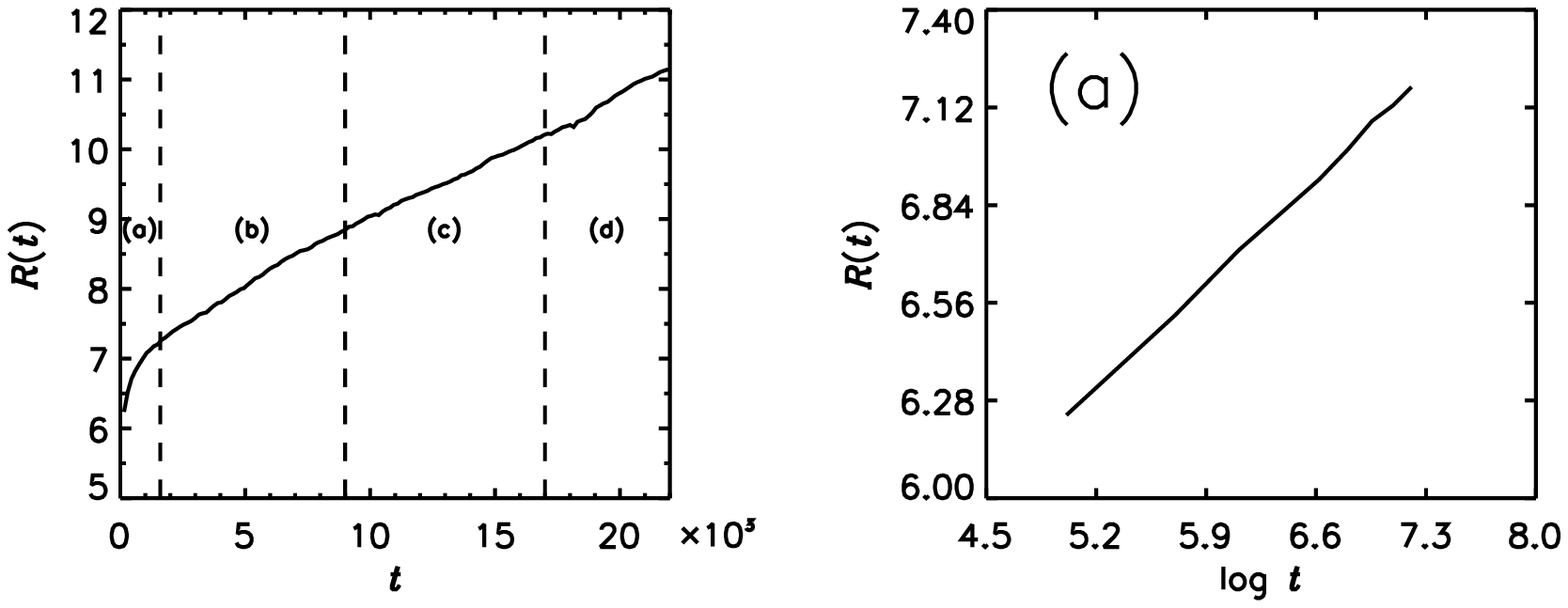, width=\textwidth}
  
  \caption{Time evolution of the characteristic domain size for the one
  dimensional case. 
  We have started from random initial conditions and used periodic boundary
  conditions. 
  Parameter values: $\eta=3.5,\ \delta=0.001$. The initial logarithmic profile
  (region (a)) becomes linear (region (c)) after a crossover (region
  (b)). The
  region (d) is related to finite size effects.}
  \label{fig:1d}
\end{center}
\end{figure}

We describe now the coarsening process that occurs when we  start from random
initial conditions for the three amplitudes.  After a short transient time, a
pattern emerges with well defined domains separated by rather
sharp kinks. Those move in such a way that neighboring kinks traveling in
opposite directions annihilate each other and, therefore, the number of domains
decreases as time increases. The final state of the system is a homogeneous
roll solution filling up the whole system.  This sequence of events happens
both for $\delta=0$ (potential case) or $\delta \ne 0$. The difference being
that  in the potential case, kinks move due to  attractive forces whereas in
the non potential regime there is an additional mechanism that brings on the
kink motion with a constant velocity given by (\ref{eq:vel}). By combining both
effects and in the case of a single domain and two varying amplitudes it is
possible to derive an equation for the rate of change in the single domain size
$R(t)$:
\begin{equation}\label{eq:size}
	\partial_{t}R(t)=2 v(\delta)-\gamma e^{-\sqrt{\eta-1}R(t)},
	\quad\gamma>0
\end{equation}
Here $v(\delta)$ is the solitary kink velocity (\ref{eq:vel}) and the second
term in the rhs is the attractive force between the kinks ($\gamma$ is a
constant independent of $\delta$ at the lowest order). When $\delta=0$ only
this second term is present.  If $\delta\neq 0$ there is a competition between
interaction and non potential effects. For the shortest times, when the kinks
are very close to each other, kink interaction will be the dominant effect
providing that $\delta$ is small enough. This leads to a growth law logarithmic
with time. As the system coarsens, the average domain size grows and the system
reaches a situation in which both effects are of the same order. Finally, when
the average domain size is large enough, the non potential kink motion will
dominate. In this regime we can consider each kink as moving at constant
velocity as a result of which the growth law is linear with time . This
prediction is supported by numerical simulations \cite{GSMT}. 
In figure \ref{fig:1d}
we show the  evolution of the average characteristic length $R(t)$ for  the
small value of $\delta=0.001$.  In accordance with the previous discussion, we
see that  the initial logarithmic growth crosses over to linear at intermediate
times. Notice that, for very late times, $R(t)$ saturates due to finite size
effects. For larger values of $\delta$, the initial logarithmic regime is very
fast because of the fast kink annihilation at the very beginning and it can
barely be observed in the simulations, which show instead a linear growth from
the very beginning. Since $\delta$ turns out to be a relevant variable for the
growth law, we expect that the scaling functions will depend on $\delta$. We
have indeed checked that the equal time  correlation function
$C(\mathbf{r},t)=\left\langle \sum_{\mathbf{x}}
A_{i}(\mathbf{x}+\mathbf{r},t)A_{i}(\mathbf{x},t) \right\rangle$ satisfies the
scaling hypothesis $C(r,t)=f(r/R(t))$, although the accuracy of the data and
the fast appearance  of finite size effects for the smaller values of $\delta$
prevent us from making a satisfactory comparison between the different scaling
functions.

\section{Two--dimensional coarsening}

The mechanisms responsible for  interface dynamics in a two dimensional system
differ from the ones holding in $d=1$. In a two dimensional potential system
whose order parameter is non conserved, interface motion is driven by
curvature. The normal velocity of a front is given by $v_{n}=-\kappa$
(Allen-Cahn law, see \cite{GSMS}),  where $\kappa$ is the local curvature of
the front line connecting two equivalent states. In the absence of other
effects, the system tends to reduce the total interface area and as consequence
it coarsens. The characteristic length grows as $R(t)\sim t^{1/2}$. 

\begin{figure}[h]
\begin{center}
 \epsfig{figure=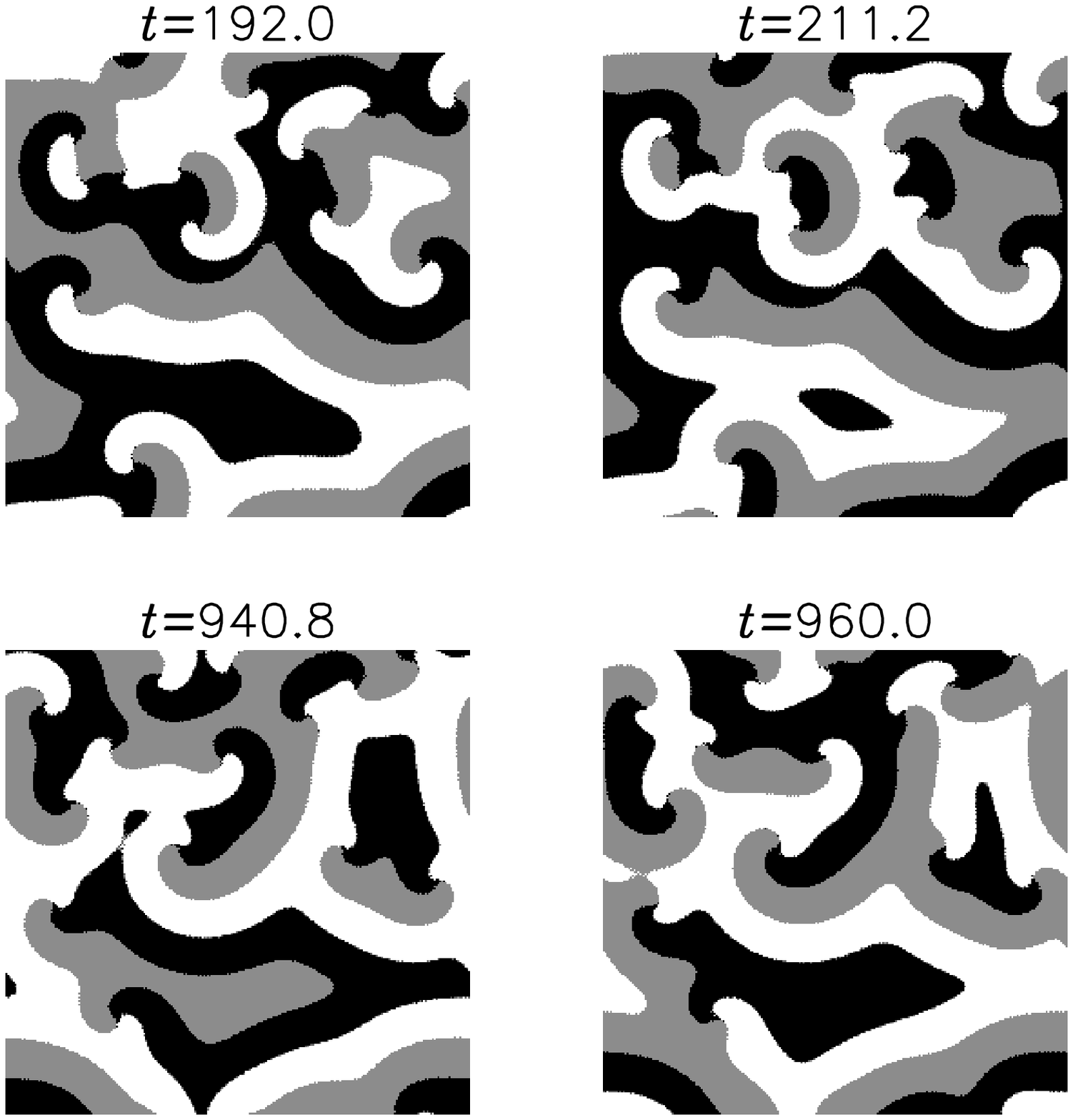,width=\textwidth}
  
  \caption{Four snapshots corresponding to the numerical simulation of the
  system (\ref{eq:modelo}) in $d=2$. 
  Parameter values: $\eta=3.5,\ \delta=0.5$. 
  The black, grey and white regions represent
  the regions occupied by $A_{1}$, $A_{2}$ and $A_{3}$ respectively.}
  \label{fig:2d}
\end{center}
\end{figure}

In our non potential system, there will be another contribution to the
interface motion coming from the fact that the interfaces connecting roll
states move at a constant velocity, namely
\begin{equation}\label{eq:vel_normal}
  v_{n}=v(\delta)-\kappa
\end{equation}
where $v(\delta)$ is the velocity of the planar front which is simply equal  to
the velocity of a one dimensional interface (eq. \ref{eq:vel}) and is zero when
$\delta$ is zero. In the case of a circular drop of radius $R$, eq.
(\ref{eq:vel_normal}) transforms into $v_{n}=-R^{-1}+  v(\delta)$. We conclude
that for a radius $R=v(\delta)^{-1} \equiv R_{c}$ the drop neither  grows nor
shrinks. Notice that this critical radius $R_{c}(\delta)$ does not appear in
the  potential problem for which a drop will always collapse in order to reduce
the surface tension \cite{MB}.  However, the most notorious fact on the
dynamical evolution in the two--dimensional system is that coarsening is 
stopped, even for values of $\delta$ smaller than the critical one or,
equivalently, when the system does not undergo the KL instability. We remind
that in the KL regime, the system does not coarsen independently of the
dimensionality. This unexpected result is due to the fact that three amplitudes
are  considered in this model. This allows the presence in the system of
``vertex points'' at which three front lines meet (see fig \ref{fig:2d}). The
role of the non potential dynamics is mainly  to rotate the front lines
around these points, preventing the system from coarsening (provided that the
system is large enough). This is  different to the situation in $d=1$ in which
the topology does not allow the meeting of three different domains.  At least
three amplitudes are necessary to stop the coarsening process. If, for
instance, one amplitude were absent, we would observe drops of one mode
immersed in a sea of the other. These drops, as in a usual nucleation theory,
either grow (spreading over the whole surface) or shrink (disappearing), and
the system coarsens.

As an evidence of the mechanisms leading to the absence of coarsening in the
two dimensional system, we present in figure \ref{fig:2d} snapshots of the
evolution of the system. We observe that the system has reached a stationary
state in which the average domain size remains approximately constant. In a
short time scale, the front lines simply rotate around the vertex points. For
larger time variations, the vertex points themselves diffuse. The exact nature
of the interactions amongst the vertices and their effect on the dynamics will
be addressed in future work.

We acknowledge financial support from DGICYT (Spain) projects numbers
PB94-1167 and PB94-1172.

\newpage


\begin{thebibliography}{10}
\bibitem{MB} In the case of a conserved order parameter, the conservation
law induces the existence of a critical radius, but this has a very
different physical origin than the one discussed in this paper.
\end{thebibliography}
\end{document}